\documentstyle[twocolumn,pra,aps]{revtex}
\draft
\newcommand{\beeq}{\begin{equation}}
\newcommand{\eneq}{\end{equation}}
\newcommand{\beeqar}{\begin{eqnarray}}
\newcommand{\eneqar}{\end{eqnarray}}
\begin{document}
\twocolumn[\hsize\textwidth\columnwidth\hsize\csname @twocolumnfalse\endcsname
\title{Cold Atoms For Testing Quantum Mechanics and Parity Violation In Gravitation}
\author{N.D. Hari Dass 
}
\address{ Institute of Mathematical Sciences, Taramani, 
Madras-600113, INDIA.}
\maketitle

\begin{abstract}
\end{abstract}

]
\section{Introduction}
Techniques of Atom trapping and laser cooling have proved to be very
important tools in probing many aspects of fundamental physics. In this 
talk I wish to present ideas on how they may used to settle certain issues 
in the foundational aspects of quantum mechanics on the one hand(sec II) and about some quantum
gravitational interactions of matter that violate parity and time-reversal,
on the other hand(sec III).

\section{Testing Quantum Mechanics of "Protective Measurements"}
A "protective" measurement is one that does not disturb the state of a
quantum system, yet it gives full information about it \cite {AAV}. This
appears to go against the general credo of  quantum mechanics which can be
encapsulated as follows:
The "value" of an observable $A$ is meaningful only in its eigenstate
\beeq
A|a_i\rangle = a_i|a_i\rangle
\eneq
If a given state $|n\rangle\ne|a_i\rangle$ for any i, the outcome of a measurement of $A$ on this state can yield any of the $a_i$ and after the measurement
that yielded some $a_i$, the system is described by the eigenstate $|a_i\rangle$.
This is called the collapse of the wave function( also called the
projection postulate of von Neumann). Further, if the state $|n\rangle$ has
the {\it unique} expansion
\beeq
|n\rangle = \sum_ic_i|a_i\rangle
\eneq
in terms of the complete basis spanned by the eigenstates of $A$, the {\it probability
interpretation } of quantum mechanics says that in a series of measurements of $A$ on an identically prepared {\it ensemble } of states $|n\rangle$, the
particular eigenvalue $a_i$ occurs with frequency proportional to $|c_i|^2$.
With this interpretation, the expectation value 
$\langle n|A|n\rangle = \sum_i|c_i|^2a_i$ of $A$ in 
$|n\rangle$ is identified with the {\it ensemble average} of the outcomes
of measurements. Note that the state $|\tilde n\rangle = \sum_i c_i
e^{i\phi_i}|a_i\rangle$, in general distinct from $|n\rangle$,also yields an
identical distribution of $a_i$. Thus to determine a state, one needs many ensemble measurements of different observables.
The number of such independent measurements needed is given by the "size" of
the density-matrix( defined by $\rho^{\dag}=\rho, tr \rho = 1, \langle O \rangle = tr \rho O$,
where $\langle O \rangle$ is  the expectation value of $O$). For example, for
a {\it pure} state($\rho = \rho^2$)  of a spin-1/2 system,this number is 2 
and for a {\it mixed} state($\rho\ne\rho^2$) of a spin-1/2 system 
this number is 3.

Based on this credo, measurements on a {\bf single} quantum state as opposed
to measurements on an {\bf ensemble} of states, has no significance in
conventional interpretations of quantum mechanics as first of all, the outcome
can be any of the eigenvalues and secondly,the state changes in an 
uncontrollable manner to the corresponding eigenstate. Repeated measurements 
subsequently will only keep
yielding the same outcome as the first measurement and the average of the
measurements bears no relation to the expectation value of $A$ in the state before measurements. As stated in the jargon of quantum mechanics, {\it a single
state has no ontological meaning}.

\subsection{A Proposal by Aharonov,Anandan and Vaidman}
In this context an extraordinary proposal was made by Aharonov, Anandan and
Vaidman \cite {AAV} who claimed that {\it when $|n\rangle$ is a non-degenerate
eigenstate of some a priori {\bf unknown} Hamiltonian, $\langle n|A|n\rangle$
can be measured for a {\bf single} quantum state {\bf without} disturbing it,
and that this can be done for arbitrary observables $A$}.

Since the original state is left undisturbed, one is free to perform a
sequence of "protective" measurements with sufficiently many different
observables and determine the single state! In other words, AAV claim to give
an {\it ontological} meaning to the state of a single quantum system. Equivalently,
they claim that the wavefunction is real and measurable even for single states
and therefore , the ensemble interpretation is unwarranted for the 
circumstances mentioned before. 

\subsection{Conventional Measurements}
To understand the results of AAV(for a detailed and critical analysis of protective measurements see Hari Dass and Qureshi \cite{qh}), it is instructive to first consider a model
for conventional measurements which as mentioned earlier lead to a collapse
of the wave function. Let us consider a quantum system described by a
Hamiltonian $H_S$ and couple the system to an apparatus described by $H_A$.
Let $Q_S$ be an
operator, corresponding to the observable of the system we wish to
measure, and let the interaction between the system and the apparatus
( in what follows, we shall use the notion of an apparatus to indicate 
a quantum system to which full information about the system can be 
transferred) be of the form

\begin{equation}
        H_I = g(t) Q_A Q_S,
\end{equation} 
where $Q_A$ is an observable of the apparatus, and $g(t)$ is the strength
of the interaction normalized such that $\int dt g(t)=1$. The interaction is
now taken to be nonzero only in a very short interval $[0, \tau]$. Let the system be in an 
initial state $|\nu\rangle$
which is not necessarily an eigenstate of $Q_S$, and let the apparatus be
in a state $|\phi(r_0)\rangle$, which is a wave packet of eigenstates
of the operator $R_A$ conjugate to $Q_A$, centered at the eigenvalue
$r_0$. The interaction $H_I$ is of such short duration, and assumed to be so
strong that the effect of the free Hamiltonians of the apparatus and
the system can be neglected. Then the combined wave function of the
system and the apparatus at the end of the interaction can be written as

\begin{equation}
        |\psi(\tau)\rangle = e^{-{i\over\hbar} Q_A Q_S } |\nu\rangle
|\phi(r_0)\rangle.
\end{equation}
If we expand $|\nu\rangle$ in the eigenstates of $Q_S$,  $|s_i\rangle$, we
obtain
\begin{equation}
|\psi(\tau)\rangle = \sum_{i} e^{-{i\over\hbar} Q_A s_i } c_i |s_i\rangle
|\phi(r_0)\rangle,
\end{equation}
where $s_i$ are the eigenvalues of $Q_S$ and $c_i$ are the expansion
coefficients. The exponential term shifts the center of the wave packet by 
$ s_i $:
\begin{equation}
|\psi(\tau)\rangle = \sum_{i} c_i |s_i\rangle |\phi(r_0+ s_i )\rangle.
\end{equation}
This is an entangled state, where the position of the wave packet gets
correlated with the eigenstates $|s_i\rangle$. Detecting the center of the 
wave packet at $r_0+s_i$ 
will throw the system into the eigenstate $|s_i\rangle$. 
\subsection{Protective Measurements}
Let us now consider the opposite
limit where the interaction of the system with the apparatus is {\it
weak} and {\em adiabatic}.  Here the system is assumed to be in a
non-degenerate eigenstate of its Hamiltonian, which is not known to the experimenter, and the interaction being
weak and adiabatic, we cannot neglect the free Hamiltonians. Let the
Hamiltonian of the combined system be

\begin{equation}
H(t) = H_A + H_S + g(t)Q_A Q_S, \label{H_full}
\end{equation}
The coupling $g(t)$ acts for a long time $T$ and goes
to zero smoothly before and after the interaction.  It is also
normalized as $\int_0^T dt g(t) = 1$. Therefore, $g(t) \approx 1/T$ is
small and constant for the  most part. If $|t=0\rangle$ is the state vector
of the combined apparatus-system just before the measurement process
begins,  the state vector after T is given by

\begin{equation}
|t=T\rangle = {\cal T} e^{-{i\over\hbar}\int_0^T H(\tau) d\tau} |t=0\rangle,
\label{psiT}
\end{equation}
where ${\cal T}$ is the time ordering operator. We divide the interval
$[0,T]$ into $N$ equal intervals $\Delta T$, so that $\Delta T = T/N$, and
because the full Hamiltonian commutes with itself at different times during
$[0,T]$,
we can write eqn(\ref{psiT}) as

\begin{equation}
|t=T\rangle = \left(exp[-{i\Delta T\over\hbar}(H_A + H_S +
{1\over T}Q_A Q_S)]\right)^N |t=0\rangle .
\end{equation}

We illustrate the special case when $Q_A$ commutes with the free
Hamiltonian of the apparatus, i.e., $[Q_A,H_A]=0$, so that we can have
eigenstates $|a_i\rangle$ such that $Q_A |a_i\rangle = a_i |a_i\rangle$
and $H_A |a_i\rangle = E_i^a |a_i\rangle$. 
Now $|a_i\rangle$ are also the exact eigenstates of the instantaneous
Hamiltonian $H(t)$, in the apparatus subspace. So, the exact
instantaneous eigenstates can be written in a factorized form
$|a_i\rangle \overline{|\mu\rangle}$ where $\overline{|\mu\rangle}$ are
system states which depend on the eigenvalue of $Q_A$, {\it i.e.}, they
are the eigenstates of ${1\over T}a_iQ_S + H_S$. Let us assume the
initial state to be a direct product of a non-degenerate eigenstate of
$H_S$, $|\nu\rangle$, and $|\phi(r_0)\rangle$:

\begin{equation}
|t=0\rangle = |\nu\rangle |\phi(r_0)\rangle .
\end{equation}
Introducing complete set of  exact eigenstates in the above equation,
the wave function at a time $T$ can now be written as

\begin{equation}
|t=T\rangle = \sum_{i,\mu} e^{{i\over\hbar}E(a_i,\mu) N\Delta T}
|a_i\rangle \overline{|\mu\rangle}\overline{\langle\mu|} 
|\nu\rangle \langle a_i| |\phi(r_0)\rangle, \label{psiT1} 
\end{equation}
where the exact instantaneous eigenvalues $E(a_i, \mu)$ can be written as

\begin{equation}
E(a_i,\mu) = E_i^a + {1\over T} \overline{\langle\mu|}Q_S\overline{|\mu\rangle}
a_i + \overline{\langle\mu|}H_S\overline{|\mu\rangle}.
\end{equation}

It should be kept in mind that the
expectation value $\langle Q_S\rangle _{\overline{\mu}}$ depends on the eigenvalue
$a_i$ of $Q_A$. The sum over $\mu$ in (\ref{psiT1}) makes it appear as
if the state is entangled. But the important point to notice is that
the basis $\overline{|\mu\rangle}$ can be made to be {\em arbitrarily}
close to the original basis, as the interaction is assumed to be weak,
so that $\overline{|\mu\rangle} = |\mu\rangle + {\cal O}(1/T) + ...$.
In the large $T$ limit, one can assume the states to be unperturbed,
and retain only terms of $O(1/T)$ in the energy (this is necessary as
$E(a_i, \mu)$ is multiplied by T in eqn(11)), which amounts to using
first order perturbation theory. This yields eigenvalues of the form

\begin{equation}
E(a_i,\mu) = E_i^a + {1\over T} \langle\mu|Q_S|\mu\rangle a_i + 
\langle\mu|H_S|\mu\rangle + O(1/T^2).
\end{equation}
In addition to this, the sum over $\mu$ disappears and only the
term where $\mu =\nu$ survives.  Thus, we can write
the apparatus part of the exponent again in the operator form

\begin{equation}
|t=T\rangle \approx e^{-{i\over\hbar}H_A T-{i\over\hbar}
Q_A\langle Q_S\rangle _\nu -{i\over\hbar}\langle H_S\rangle _\nu T}
|\nu\rangle |\phi(r_0)\rangle .
\end{equation}

Now, it is easy to see that the second term in the exponent will shift
the center of the wave packet $|\phi(r_0)\rangle$ by an amount $
 \langle\nu|Q_S|\nu\rangle$:

\begin{equation}
|\psi(T)\rangle = e^{-{i\over\hbar}H_A T-{i\over\hbar}\nu T}
|\nu\rangle |\phi(r_0+\langle Q_S\rangle _\nu)\rangle .\nonumber\\
\end{equation}
This shows that at the end of the interaction, the center of the
wave packet $|\phi(r_0)\rangle$ shifts by $\langle\nu |Q_S|\nu\rangle$.

The idea behind this approximation is that in $\overline{\langle
\mu}|\nu\rangle$ only one term is large and close to unity, and rest of
the terms are very small, of the order $1/T$. Making $T$ very large,
one can make the smaller terms arbitrarily close to zero. Thus, the
state is effectively not entangled, and so the original wave function is
not destroyed during the measurement.  Looking at the position of the
wave packet, one can determine the expectation value  $\langle
Q_S\rangle _\nu $. This, basically, is the essence of 
protective measurements.
\subsection{Protective Measurements Do Not Work For Single Systems}
As we saw in the last section,complete disentanglement of the apparatus
and system wavefunctions only happens in the $T\rightarrow\infty$-limit,which
though mathematically well defined, can never be realised physically.Of course
the degree of entanglement can be made arbitrarily small in any physical 
realisation,but never {\bf zero}.In the case of {\it ensemble} measurements,a tiny
admixture in wavefunctions is of no consequence as it will only change the
probability distribution by a very small amount.

But for a single system,even an extremely {\it tiny} entanglement with the
apparatus can have a disastrous effect!This is because for single systems 
the outcome of the first measurement can be anything and not necessarily the
outcome with the largest probability as calculated from the probability
interpretation of quantum mechanics.In particular,the first outcome could
throw the system into a state quite far from the initial state.Then one can
not rely on successive protective measurements to yoeld any information 
on the original state.Stated differently,"protective" measurements on single
states can never be done with 100\% confidence.Hence the {\it wavefunction
can not be deemed to be real}.
\subsection{A Pragmatic Interpretation}
Though the idea of a protective measurement does not work for single systems
and hence fails to give any reality to the wavefunction,from a practical
point of view the proposal has some very interesting possibilities.When
performed on an {\it ensemble},the effect of the tiny entanglement becomes
negligible and this means that the system ensemble which was {\it pure} to
begin with,remains {\it pure} to a high degree even after the measurement!
This is in contrast to the conventional measurement where a pure state
density matrix $\rho_{in} = |n\rangle\langle n|
=\sum_{ij}c_i^*c_j|a_i\rangle\langle a_j|$ turns into the mixed state
density matrix $\rho_{fin}=\sum_i|c_i|^2|a_i\rangle\langle a_i|$ after
measurement.This could have a lot of practical applications.Also,since the
ideal protective measurement allows one to straightaway measure expectation
values of arbitrary observables,in practice much smaller ensembles could yield
accuracies comparable to what one obtains in conventional measurements.It is
worth the while to experimentally test some of these consequences.In the next
section we show, through a simple example, how cold atoms could be used for
this purpose.
\subsection{Cold Atoms for Testing Protective Measurements:An Example}
Consider the set-up for a typical  Stern-Gerlach experiment represented by
the Hamiltonian 
\beeq
H ={{\vec P}^2\over 2M} -\mu B_0\vec\sigma\cdot\vec {\tilde n}-\mu g(t) 
{\tilde B}_i x\vec\sigma\cdot\vec n.
\eneq
Here M is the mass of ,say,the silver atom.
As before, $g(t)$ is taken to be ${1\over T}$. {\em It should be noted
that $B_0\vec {\tilde n}$ is an a priori unknown magnetic field}. Consequently
we shall not vary $B_0\vec {\tilde n}$ during the experiment. The initial state
is chosen to be

\beeq
|t=0\rangle = e^{ip_0y} |\tilde +\rangle ;~~~~~~~\vec\sigma\cdot\vec{\tilde n}
|\tilde \pm\rangle=\pm|\tilde \pm\rangle .
\eneq
It should be emphasized that this initial state is a priori unknown.We first
treat this case by ignoring the kinetic term for the silver atoms which is
tantamount to assuming them to be heavy.Then,
the Hamiltonian of eqn(16) is the Hamiltonian of the spin-1/2 particle
in the effective magnetic field

\beeq
\vec B=B_0\vec{\tilde n}+{\tilde B}_i {x\over T}\vec n ,
\eneq
whose eigenstates are given by
\beeq
H|\pm\rangle=\pm \mu B|\pm\rangle .
\eneq
Consequently, the state at $t=T$ is given by
\beeq
|t=T\rangle=[\cos {\theta\over 2} e^{i\mu BT} |+\rangle+\sin{\theta\over 2} 
e^{-i\mu BT} |-\rangle ,
\eneq
where $\theta$ is the angle between $\vec B$ and $\vec{\tilde n}$. As $T\rightarrow \infty$,
$\theta\rightarrow 0$ and $|+\rangle\rightarrow|\tilde +\rangle$. Also
\beeq
B\rightarrow B_0+{\tilde B}_i{x\over T}\vec n\cdot\vec{\tilde n}.
\eneq
Thus
\beeq
|t=T\rangle  \rightarrow 
e^{i\mu B_0T} e^{i(p_0y+\mu {\tilde B}_i
\vec n\cdot\vec{\tilde n}x)} |\tilde +\rangle .
\eneq
Hence the momentum of the apparatus in the x-direction shifts by $\mu {\tilde B}_i\vec
n\cdot\vec{\tilde n}= \langle\mu {\tilde B}_i\vec\sigma\cdot\vec
n\rangle_{\tilde +}$, while the system remains in the same state to
begin with.Now we incorporate the essential effect of the atom kinetic 
term which was ignored;this is to make the initial wave packet spread in
a manner that is familiar from free particle quantum mechanics.So 
, we take the initial wavefunction to be a plane wave in the y-direction
but a wavepacket
$\phi(\epsilon)$ centred around the origin and of width $\epsilon$ in the x,z-directions.Thus the
wave function after the adiabatic interaction is
\beeq
|t=T\rangle  = 
e^{i\mu B_0T} e^{ip_0y}e^{i\mu {\tilde B}_i
\vec n\cdot\vec{\tilde n}x} |\tilde +\rangle .
|\phi(\epsilon(T))\rangle 
\eneq
where

\beeq
\epsilon(T)^2= {1\over 2}(\epsilon^2+{T^2\over M^2\epsilon^2})
\eneq
\subsection{A Possible Experimental Set Up}
Let us see if the previous example can be experimentally realised.We could
consider a beam of cold atoms of width $\epsilon=0.1~cm$.In addition to the
various conditions of weakness and adiabaticity of interactions,we have 
to ensure that the momentum due to the uncertainty $\epsilon$ is much less than
the momentum picked up due to the measurement interaction.Otherwise, the spreading
of the wavepacket will completely obliterate the effects of the measurement.
Let us take $B_0=1~ Gauss$.Let the beam move in the y-direction and let the
spatial extent of the inhomogeneous magnetic field in the x-direction be 
$L= 30~ cms$. Let the velocity of the cold atoms be $v$.Now we define a "critical
velocity " $v_c$ by
\beeq
\mu B_0 {L\over v_c} = 1
\eneq
For $\mu = 1$ nuclear magneton,we get $v_c = 3\times 10^5~ cm/sec$.The
relation between the inhomogeneous field $B_i$ in the laboratory and ${\tilde
B}_i$ introduced in the last section is $B_i={\tilde B_i}{v\over L}$.This
can be written slightly differently as
\beeq
B_i = {v\over v_c}3{\tilde B_i}\times 10^{-7} Gauss ~cm^{-1}
\eneq
The condition of weakness of the inhomogeneous Stern-Gerlach field i.e
$B_0 >> B_i x_{max}$ becomes,on taking $x_{max}=1~~cm$
\beeq
1 G >> {v\over v_c}3{\tilde B_i}\times 10^{-7} G
\eneq
If velocities as low as 1 cm/sec can be achieved for the atoms,${\tilde B_i}$
can taken to be $10^{11}$ giving a maximum inhomogeneous field strength of
$\simeq 0.1 G$.Now the momentum imparted to the beam in the x-direction by
the measurement interaction is
\beeq
P_{meas} = {\mu L\over v_c}3 {\tilde B_i} 10^{-7} G cm^{-1}
\eneq
With the parameters we have chosen this is $3{\tilde B_i} 10^{-7} cm^{-1}$(we
are now using the natural units $\hbar=c=1$).With ${\tilde B_i}=10^{11}$ we get
$P_{meas} = 3\times 10^4 cm^{-1}$. As the momentum due to beam width 
uncertainty ${1\over \epsilon}$ is 10 $Cm^{-1}$,we see that the momentum
imparted by interaction is much larger so the effect of the sprading of the wavepacket is negligible.If we take the mass of the atom to be 50 amu,the velocity
imparted is $6\times 10^{-2}$ cms/sec.If after the interaction,the atoms are
allowed to stream freely for about 30 secs, the displacement is about 2 cms,
much larger than the beam width.

Then we can use many of the known techniques to establish that most of the 
initial ensemble is in tact and at the same time we would have measured the
expectation value of an observable (in this case $\vec S\cdot \vec n $).
The choice of parameters is only indicative and a more careful study is needed
to optimise them.
\section{Testing Parity Violating Quantum Gravitational Interactions}
General Theory of Relativity(GR) is extraordinarily succesful in accounting
for almost all classical gravitational phenomena so far. This prompts one to
asak whether this success implies the correctness of GR at the quantum level
too. I'll outline here how cold atom techniques can shed light on some aspects
of this fundamental question.

The crux of GR is the equivalence principle (for the moment let us not bother
to distinguish between the so called "strong" and "weak" versions of this),
which asserts the local equivalence of the {\bf dynamical} effects of gravitation
and the {\bf kinematical} effects of an accelerated frame. Stated
otherwise,an observer freely falling in a gravitational field would,according 
to GR,feel that he is in an inertial frame and to that observer all laws of 
nature would look like laws in empty space.

To appreciate the rest of my talk, it is worthwhile elucidating what an
inertial frame is. One way of characterising inertial frames is to say they
are the frames in which a stationary source of light sends out spherical
wavefronts i.e light propagates isotropically with constant velocity c.

The other characterisation of an inertial frame is to say they are the frames
in which an isolated gyroscope,for example,maintains its total angular
momentum $\vec L$.

We shall call the latter a "rotational" characterstic of inertial frames.The
two characterisations are logically independent.One knows that momentum 
$\vec P$ and angular momentum $\vec J$ are independent generators of the Lorentz
group( a group of relevance when the gravitational fields are weak).  

So a natural question to ask is how does the gravitational field couple to
the different generators of the Lorentz group.It is well known that the
generator $ \vec J$ can be split into an orbital part and an intrinsic
part.Under classical circumstances i.e when the macroscopic averages of
$\vec J_{intrinsic}$ is zero,only the orbital part is relevant and this
is not independent of $ \vec P$ in the sense that $\vec L = \vec R \times
\vec P$.In particular,knowing how point particle momenta couple to a 
gravitational field and modelling a gyroscope as a collection of point
particles in rotation,one can work out the coupling of the "spin angular
momentum" of the gyroscope to the gravitational field.It can be shown that
this is indeed in accordance with GR.

Thus the precession of gyroscopes in a gravitational field is not {\it
a logically independent test} of GR.On the other hand,with the advent of
quantum mechanics and quantum field theory, we know that intrinsic angular
momenta can not be consistently modelled by extended rotating objects. 

According to GR,the intrinsic spins must respond the same way as classical
orbital angular momenta to gravitational fields.Otherwise, in a freely
falling frame there would be differential precessions and that would not 
look like flat space physics! {\bf But do they?}.

Unfortunately there is no experimental evidence for this mainly because for
all objects with which classical GR tests are carried out,the intrinsic
spins are thermodynamically averaged out viz star light, pulsars in binary
systems etc.

It is the purpose of my talk here to argue that trapped atom 
and laser cooling techniques have reached a stage where we can begin to
experimentally tackle these fundamental questions.It should be remarked that 
these new tests should not be confused with other tests like interferometric
tests which have only established the quantum manifestations of the
{\it Newtonian Potential}.

Before discussing the actual experiments,let us write down the most general
parametrisation of the leading order long range spin-dependent potential
\cite {ndh}:
\beeq
V(\vec r) = \alpha {GM\over cr^3}\vec S\cdot \vec r + \beta {GM\over c^2r^2}\vec S\cdot\vec v+\gamma {GM\over c^2r^3}\vec S\cdot (\vec r\times\vec v)
\eneq
Leitner and Okubo \cite{leit} had parametrised these discrete symmetry 
violating potentials as
\beeq
V(\vec r) = V_0(r)\{1 +A_1\sigma\cdot\hat r+{A_2\over c}\sigma\cdot\vec v+
{A_3\over c}\vec\sigma\cdot(\hat r\times\vec v\}
\eneq
where$V_0(r)$ is the usual Newtonian potential.As this parametrisation has
generated a lot of confusion,we wish to remark on it. While this 
is fine as a phenomenological parametrisation for spin-1/2 particles,it
lacks a certain universality in the sense that $\vec \sigma$ are the roation
generators in the spin-1/2 representation only.Of course one could argue that
for the general case $\vec \sigma$ should be replaced by $\vec S/\hbar$ but
this is unnatural because of the explicit appearance of an inverse power of
$\hbar$.Our parametrisation avoids these problems.It can also be brought
into the Leitner-Okubo form but with $A_i$ no longer {\it constants} but
depending on $r$.Thus limits on $A_i$ will be {\it context dependent}!In fact
for massive particles $A_1=\alpha{\hbar\over mcr}$ etc.

GR predicts that $\alpha=0,\beta=0,\gamma=2$.
Any deviation from these values
implies a breakdown of the equivalence principle and as shown in \cite{ann}
actually means even a breakdown of the very structure of GR.In particular,
they imply violation of the equivalence principle as well as a violation of 
Local Lorentz Invariance(LLI),which is really the true symmetry content of GR. That
these effects violate these hallowed principles of GR can be expressed in a
visually striking manner by noting that even in the freely falling frame 
elementary particle spins would be precessing and hence not all effects of
gravitation are equivalent,even locally,to the effects of being in an accelerated frame! They also imply that discrete symmetries like P,C,T are violated in
gravitational interactions.In \cite {ann} I have presented a very detailed
analysis of the significance of $\alpha\ne 0,\beta\ne 0$ and the severe
restrictions such a circumstance would place on many theories of gravitation 
like GR and its variants, Einstein-Cartan theories,non-symmetric theories of
gravitation,Poincare-gauge theories etc.(It should be noted that the non-symmetric theories I have in mind are those such that the antisymmetric part of the "metric" has its source only in quantum effects).Thus it is
extremely important to settle the status of these parameters experimentally.

The high precision tests of GR shed no light on these parameters as intrinsic
spins are thermodynamically averaged to zero in such tests. One will have to
perform gravitational experiments with spin-polarised objects.The experiments
proposed by me are of that Genre.

Of the three effects,all terms are of comparable magnitude for photons.For
massive particles,the $\alpha$ -term is $c/v$ times bigger than the 
other terms.The main observable consequences of the $\alpha-$term are:

{\bf Differential Acceleration}:\\
\beeq
{a_+-a_-\over a} = 2\alpha{\hbar\over mcR}
\eneq
where $a_\pm$ refer to the acceleration of particles with spin polarised 
parallel(anti-parallel) to the local gravitational field and R is the 
radius of the earth.In magnitude this works out to $\simeq 10^{-22},
10^{-19},10^{-23}$
for neutrons,electrons and atoms respectively.

\noindent
{\bf Spin Precession}:\\
The spin-vector $\vec S$ will precess around the local gravitational field
according to
\beeq
{d\vec S\over dt} =\alpha {GM\over cR^3} \vec R\times\vec S
\eneq
This effect is independent of the mass of the particle and the precession rate
is $\simeq 4.5~\alpha $nHz.

\noindent
{\bf Energy Difference between polarised particles}:\\
Let us consider spin-1/2 particles.The energy difference between particles
with spin along the direction of the gravitational field and those with spin
anti-parallel to the local gravitational fields is given by:
\beeq
\Delta E ={GM\over cR^2}\hbar\alpha
\eneq
Numerically this is $\simeq 3~\alpha 10^{-23}eV$.\\

\noindent
{\bf Existing Limits}\\
As mentioned earlier,the existence of these effects would imply  differential
acceleration of particles with spin along and opposite the local gravitational
field.Unfortunately such experiments have not been conducted yet.However,by
ascribing the current uncertainties in Eotvos type experiments to these 
effects one could arrive at the limit $\alpha < 10^9$ which is not a very useful
bound. It would be instructive to perform "fifth-force" type experiments
with test objects where a substantial fraction of the intrinsic angular
momentum of the body is due to the intrinsic spin-angular momentum of the composites(as
opposed to the quantised orbital angular momentum).

Leitner and Okubo estimated $A_i$ by ascribing the uncertainties in the 
hyperfine splitting of hydrogen arising out of an uncertainty in the fine structure
constant(at that time about 1ppm) to possible non-electromagnetic phenomena.
And ascribing them fully 
to the discrete symmetry violating potential they concluded $A_i< 10^{-11}$.
While this sounds impressive,it implies $\alpha<10^{10}$ which is far from 
impressive!Currently,the fine structure constant is known to about .05 ppm
but the discrepancy between theory and experiment is still at the level of
0.5 ppm \cite {ramsey}. 

In fact the most conservative limit on $\alpha$ from the Hfs in hydrogen can
be obtained by ascribing the uncertainty in measurement to the extra effects
considered here.This error is about 1 mHz\cite {ramsey}.This translates 
to a limit of $\alpha < 200$.

In fact a good way of estimating these effects is by looking for differential
bending of polarised electromagnetic waves at the solar limb.In fact such an
effect was looked for by Harwit et al \cite{harwit} who,with the 
parametrisation
of the potential given by
\beeq
V_{eff} = \pm\alpha^\prime V_0(r)
\eneq
where $\pm$ refers to the two states of circular polarisation.The relation 
to our parametrisation is $\alpha^\prime=\alpha{\lambda\over 2r}$ where 
$\lambda$ is the wavelength.Harwit et al found $\alpha^\prime \simeq 10^{-2}$
with 13 cm radio waves at the solar limb.Later Dennison \cite {dennis} obtained the more
accurate value of $\alpha^\prime\simeq 10^{-6}$ which translates to 
$\alpha\simeq 10^4$. As these experiments were done long ago,perhaps current
state of the art may already give a chance to probe $\alpha\simeq 1$.
\subsection{Astrophysical Constraints} 
There have been several attempts to see if the parameters of discrete symmetry violations in gravitation could be constrained using astrophysical data.
Almeida et al \cite{almeida} used the limits on differential propagation
of neutrinoes and photons to set the limits $|A_i|< 10^{-3}$ where $A_i$
are the parametrisations used by Leitner and Okubo.As remarked already,these are context dependent and their relation to $\alpha,\beta$ in this context
is $A_1=\alpha{\hbar c\over 4Eb},A_2 = \beta{\hbar c\over 4Eb}$ where $b$
is the impact parameter relative to the galaxy and $E$ the energy of the
particles.Their limit translates to $\alpha < 10^{30}$, not a very useful
bound.

Losecco et al \cite {losecco} tried to use the observed pulse-widths of pulsars to 
constrain the differential propagation times of polarised photons and
obtained $A_1 < 2\times 10^{-10},A_2 < 6\times 10^{-11}$.They thought
that their limits on $A_1$ were poorer than what \cite {leit} obtained,
but again due to context dependence the two are not related.This translates to
$\alpha < 10^{23}$.

Choudhury et al \cite{cqg} tried to estimate these parameters by looking
at their contribution to helicity flip scattering of massive neutrinoes
and their consequent effect on the cooling of neutron stars.Assuming
a mass of 1 KeV for $\tau$-neutrinoes,they could constrain $\alpha$
to be less than 300.

Nodland and Ralston \cite {nodland} report that their is a systematic 
cosmological effect that rotates the polarisation of photons.It is
very important to see whether such an effect can be explained on the 
basis of our model.There is also a recent work of Lue et al \cite {cmbr}
which has analysed some very specific models with P violation for their
cosmological signatures. We plan to investigate these signatures from the
point of view of our model.There is also considerable theoretical interest
in P violations from the point of view of a variety of models \cite {mukho},
as well as for their implications for global space-time issues \cite{jeeva}.
\section{Atom Techniques}
\subsection{Mercury Cell}
In the experiment of Venema et al \cite {mercury} two ground state Hg isotopes are trapped in the same cell.The electronic con
figuration being $^1S_0$,the two isotopes are fully "nuclear spin polarised".
The two isotopes under consideration are $^{199}Hg(I=1/2)$ and 
$^{201}Hg(I=3/2)$.

One of the most problematic backgrounds for experiments looking for small
spin precession rates like the Ramsey type Electric Dipole Moment 
measurements or our proposed experiments to look for discrete symmetry
violations in gravitation are the very tiny stray magnetic fields that 
mimic the effects we are trying to study.For the gravitational experiments
even stray fields as small as $10^{-11}$ Gauss can simulate an effect
equivalent to $\alpha\simeq 1$. 

The advantage of trapping two isotopes in the same cell is that they will
both experience nearly the same stray magnetic field. The effect of the stray
magnetic field can then be eliminated through an appropriate combination of observables. In the actual set up ambient magnetic fields were reduced to the level
of $< 20 \mu G$.In addition an uniform field of less than $10 mG$ was also
applied in the z-direction.The direction of this field was flipped every
hour.The Hamiltonian for the system can be cast as
\beeq
H= -g_I\mu_N\vec I\cdot\vec B+{\cal A}\hat I\cdot\hat r
\eneq
where $\vec I$ is the nuclear spin,$\vec B$ the applied uniform field,$\mu_N,{\cal A}$ the nuclear gyromagnetic ratio and the quantity parametrising the
additional effects one is looking for,respectively.The effect of the stray 
magnetic field can be taken into account by changing the Hamiltonian to
\beeq
H_{stray} = H - g_I\mu_N\vec I\cdot\vec B_{ran}
\eneq
Then the observed frequency of the NMR-line is given by
\beeqar
\nu_{199} & = & -g_{199}\mu_N(B+B_{ran}<\cos \theta_{ran}>)+{\cal A}_{199}\cos \phi\nonumber\\
\nu_{201} & = & -g_{201}\mu_N(B+B_{ran}<\cos \theta_{ran}>)+{\cal A}_{201}\cos \phi
\eneqar
where $\phi,\theta_{ran}$ are the angles between the direction of the uniform
magnetic field and the local gravitational field,random magnetic field
respectively.In general $\theta_{ran}$ is time-dependent but we assume that
it is constant over the time scale of $g_I\mu_NB$.Now there are two observables
(related to each other) that are insensitive to the stray magnetic field;these
are i) ${\cal R}=\nu_{199}/\nu_{201}$ and 
ii) ${\cal S}={\nu_{199}\over g_{199}} - {\nu_{201}\over g_{201}}$,
which are given by:
\beeqar
{\cal R} &=& {g_{199}\over g_{201}}\{1+{1\over \mu_N B}
({{\cal A}_{201} \over g_{201}}-{{\cal A}_{199}\over g_{199}})\cos \phi\}
\nonumber\\
{\cal S} &= &
({{\cal A}_{201} \over g_{201}}-{{\cal A}_{199}\over g_{199}})\cos \phi
\eneqar
For the gravitational context,${\cal A} = \alpha{GM\over R^2c} I$.Hence for this case
\beeq
{\cal S} = \alpha({GM\over R^2c}){\cos \phi\over 2}({3\over g_{201}}-
{1\over g_{199}})
\eneq
Thus this method is not suitable when the gyromagnetic ratios of the
isotopes are in proportion to their nuclear spins.

It is remarkable that the measurements are so accurate that proper account
of earth's rotation should be taken into account! The change in the observed frequency is $\nu\rightarrow\nu\pm\Omega_E\cos\theta$ where $\Omega_E$ is the
earth's angular velocity and $\theta$ the angle between the earth's axis
of rotation and the z-direction. Numerically $\Omega_E\simeq 11.6 \mu Hz$. It
is best to choose $\theta$ small as the dependence on it would be like 
$\theta^2$.This will help minimise errors due to mechanical misalignment.
The systematic error due to this was found to be $\Delta={\cal R}_+
-{\cal R}_- \simeq 10^{-8}/\nu(Hz)$.

The quoted result for ${\cal A}$ is
\beeq
{\cal A} = -0.12\pm0.14\pm0.15(Syst) \mu Hz
\eneq
In eV this translates to ${\cal A} \simeq 2.2\times 10^{-21} eV$.Consequently
the implied limit on $\alpha$ is $\alpha\simeq 70$.

Thus the accuracies of this experiment need to be pushed just a few orders of 
magnitude more before we can begin to probe $\alpha\simeq 1$.

I list here some possibilities for future. Instead of Hg Vapour cell, one
could imagine trapping two different isotopes in an atom. In discussions
with Y. Takahashi at Kyoto University,the use of a Magneto-Optic Trap(MOT)
has been identified.Here it does not seem too idealistic to expect to
achieve the following parameters: Number of trapped atoms $\simeq 10^8$,
dipole force time $\simeq 10 sec$,total measurement time $\simeq 10^6 secs$
leading to statistical errors $\simeq \mu Hz/\sqrt N$.Systematic errors
are hard to estimate as yet.By also studying the full $\phi$-dependence,
one may hope to get some control over the systematic errors. With all
these features,it is hoped that we may begin to probe the $\alpha\simeq 1$
region.
\subsection{Atoms in Traps}
Another class of experiments that can be very succesful in putting useful
limits on parameters like $\alpha$ are experiments where ions are trapped 
in a Penning trap.In the experimental set up of Weinland et al \cite{wine},about 5000
atomic $^9Be^+$ in its ground state characterised by $^2S_{1/2}$ are
trapped in a Penning trap with magnetic field $B_0 \simeq 0.8194 $ Tesla.
The hyperfine structure of the ground state is characetrised by $F=1$
where $\vec F = \vec J(electron)+\vec I(nuclear)$.The hyperfine transition 
that was driven was $(F=1,m_F=0)\rightarrow(F=1,m_F=-1)$.

The ions were trapped and laser cooled to reduce doppler shifts.First,they
are optically pumped to $m_F=0$ using a  combination of laser and rf
coils.Then the required transition($m_F=0\rightarrow m_f=-1$) was driven
via rf.Finally the changes in the population of the two states were
monitored by the changes in the scattered laser light.Changes in $\nu_0$
were looked for as the magnetic field $B_0$ was flipped.Again,the effect of the 
earth's rotation has to be subtracted.The final result is
\beeq
\Delta \nu_0 \simeq -6.4\pm2.9\pm6.4 \mu Hz
\eneq
where the main source of error of $6.4 \mu Hz$ is due to the pressure shift
variation when the magnetic field is flipped.Using quadrature,one can put
the limit
\beeq
\Delta \nu_0 < 13.4 \mu Hz
\eneq
which translates to the limit
\beeq
\alpha < 300
\eneq
The authors of this experiment claim that the error due to pressure shift 
variation can be minimised by using cryogenic pumping.Also,the number of stored
ions could be increased to $10^7$.The time to drive the $\nu_0$ transition 
could also be increased to 100 secs.With all these improvements the limits
on $\Delta \nu_0$ could be improved to $\simeq 3 nHz$ which would push the
limits on $\alpha$ to $\simeq 1$.
\subsection{Experiments looking for Local Lorentz Invriance Violations}
In the early sixties the so-called Hughes-Drever experiments  
looked  for violations of Local Lorentz Invariance  which is at the heart
of GR.These experiments looked for a particular manifestation of the violation
of LLI viz. dependence of the inertia of a body on its velocity and
orientation relative to some preferred frame.While the early limits on
${\delta m\over m}$ were $\simeq 10^{-23}$, these have been improved 
substantially in
recent times. Lamoreaux et al \cite{lamor1} improved this to ${\delta m\over m}
\simeq 10^{-28}$ or equivalently to splittings of the order of 500 nHz. Chupp
et al \cite{chupp} also obtained comparable limits of 450 nHz
($2\times 10^{-21} eV$). Most recently Berglund et al \cite {berg} have obtained limits of 110 nHz
for nucleons.

It is tempting to see if one could use these experiments to limit our $\alpha$
parameter also. It should be recalled that one of the consequences of the 
existence of $\alpha$ is indeed violation of the equivalence principle and 
of LLI.One
could imagine the earth's gravitational  field providing the preferred 
direction.

In all these experiments nuclei are polarised and allowed to freely precess
about the direction of a magnetic field which is fixed in the earth frame.As
the earth rotates,the orientation of this magnetic field with respect to 
some preferred direction keeps changing with a 24 hr cycle(nearly) and these
experiments look for changes in the Zeeman level splittings that are correlated
with this cycle.Unfortunately,for testing our modcel these experimental set ups
{\it are not adequate} as the angle between the magnetic field and the earth's
gravitational field does not change.Nevertheless the extreme sensitivities
achieved in these experiments suggests readopting them to look for the type of
effects that we have been discussing.

This underscores the fact that these experiments are looking at very
specific models of LLI and are not looking at generic LLI violations.
In some sense they are looking for kinematical effects incorporating
LLI violations whereas our effects embody LLI violating dynamics.A
wide class of LLI violating models have recently been analysed by 
Glashow et al \cite{glashow}.

I am thankful to Y. Takahashi, H. Funahashi and Akira Masaike of Kyoto
University for many illuminating discussions on the feasibility of improving
the limits on the parity violating parameters, and for Y. Takahashi for
many discussions on realising protective measurements.

\end{document}